\documentclass[conference]{IEEEtran}
\IEEEoverridecommandlockouts

\usepackage{amsmath,amssymb}
\usepackage{graphicx}
\usepackage{cite}
\usepackage{booktabs}
\usepackage{tikz}
\usetikzlibrary{arrows.meta,positioning}
\usepackage{pgfplots}
\pgfplotsset{compat=1.18}
\usepgfplotslibrary{groupplots}
\definecolor{cblind}{HTML}{2A78D6}
\definecolor{cgenie}{HTML}{EB6834}
\definecolor{cpilot}{HTML}{1BAF7A}
\definecolor{cnoncoh}{HTML}{EDA100}
\definecolor{cpit}{HTML}{E87BA4}
\definecolor{ctheory}{gray}{0.2}
\tikzset{v3panel/.style={font=\footnotesize, inner sep=1.5pt}}
\pgfplotsset{
  v3axis/.style={
    scale only axis, width=0.84\columnwidth,
    tick label style={font=\scriptsize}, label style={font=\footnotesize},
    legend style={font=\scriptsize, draw=black!60, line width=0.4pt, fill=white, fill opacity=0.9,
      text opacity=1, inner sep=1pt, row sep=-1.5pt, cells={anchor=west}},
    legend cell align=left,
    grid=major, major grid style={black!12, line width=0.3pt},
    axis line style={line width=0.5pt}, every tick/.style={black, line width=0.4pt},
    tick align=inside, xtick pos=bottom, ytick pos=left, xlabel style={yshift=2pt}, ylabel style={yshift=-2pt},
    every axis plot/.append style={line join=round},
    clip marker paths=false, clip bounding box=upper bound,
  },
  v3th/.style={#1, line width=0.9pt, no markers, forget plot},
  v3mc/.style={#1, only marks, mark options={fill=white, line width=0.6pt, solid}, mark size=1.5pt,
    line width=0.6pt, forget plot,
    error bars/.cd, y dir=both, y explicit, error bar style={line width=0.5pt, solid},
    error mark options={rotate=90, mark size=1pt, line width=0.5pt}},
  v3leg/.style={#1, line width=0.9pt, mark options={fill=white, line width=0.6pt, solid}, mark size=1.7pt},
  mgenie/.style={color=cgenie, mark=square*},
  mpilot/.style={color=cpilot, mark=triangle*},
  mblind/.style={color=cblind, mark=*},
  mnoncoh/.style={color=cnoncoh, mark=diamond*},
  mpit/.style={color=cpit, mark=triangle*, mark options={rotate=180}},
}

\usepackage{microtype}
\usepackage{xcolor}
\usepackage{textcomp}
\usepackage{orcidlink}

\usepackage[acronym,nomain]{glossaries}
\newacronym{ceofdm}{CE-OFDM}{constant-envelope orthogonal frequency-division multiplexing}
\newacronym{ofdm}{OFDM}{orthogonal frequency-division multiplexing}
\newacronym{pit}{PIT}{phase-domain injected training}
\newacronym{zc}{ZC}{Zadoff--Chu}
\newacronym{st}{ST}{superimposed training}
\newacronym{dft}{DFT}{discrete Fourier transform}
\newacronym{papr}{PAPR}{peak-to-average power ratio}
\newacronym{qam}{QAM}{quadrature amplitude modulation}
\newacronym{lfm}{LFM}{linear frequency-modulated}
\newacronym{nlfm}{NLFM}{nonlinear frequency-modulated}
\newacronym{cp}{CP}{cyclic prefix}
\newacronym{snr}{SNR}{signal-to-noise ratio}
\newacronym{pfr}{PFR}{peak-to-floor ratio}
\newacronym{dtr}{DTR}{direct-to-target ratio}
\newacronym{fft}{FFT}{fast Fourier transform}
\newacronym{dc}{DC}{direct current}

\hypersetup{hidelinks}

\title{Payload-Blind Bistatic Delay--Doppler Sensing with Phase-Injected Chirps in CE-OFDM}

\author{%
\IEEEauthorblockN{%
Gustavo P. Gon\c{c}alves$^\dagger$\,\orcidlink{0009-0000-8260-4390},
Henrique L. Senger$^\dagger$\,\orcidlink{0009-0004-1586-8168},
Bruno S. Chang$^\dagger$\,\orcidlink{0000-0003-0232-7640}
 and Giuseppe Thadeu Freitas de Abreu$^\ddagger$\,\orcidlink{0000-0002-5018-8174}
}
\IEEEauthorblockA{%
$^\dagger$CPGEI/Electronics Department,
Federal University of Technology -- Paran\'a, Curitiba, Brazil\\[0.15ex]
$^\ddagger$School of Computer Science and Engineering,
Constructor University Bremen, Germany\\[0.15ex]
Emails: \{gustavog.1999, hsenger\}@alunos.utfpr.edu.br, bschang@utfpr.edu.br,  gabreu@constructor.university%
}
}

\begin{document}

\maketitle

\begin{abstract}
%CE-OFDM enables power-efficient transmission, but its Hermitian-symmetric structure makes dedicated sensing pilots costly, and pilots cannot be superimposed without destroying the constant envelope. We show that injecting a known chirp, here a ZC sequence, into the CE-OFDM phase enables payload-blind bistatic delay--Doppler sensing without pilot frames or a power split: the residual carrier, i.e., the nonzero mean of every CE-OFDM signal, becomes an implicit superimposed pilot for a receive-only node. A Doppler-aligned receiver integrates this reference coherently across frames while the payload averages out. In contrast, the frame averaging of PIT, designed for quasi-static channel estimation, suffers from target Doppler, losing up to 32~dB for the simulated target with 16 averaged frames. We derive the payload-induced floor, detection probability and direct-path tolerance in closed form; the tolerance grows with the number of integrated frames. Payload-blind sensing costs at most 2.8~dB relative to a payload-aware genie at the simulated modulation index and matches a dedicated-pilot benchmark of equal reference energy at low SNR, without its 53\% frame overhead.
\Gls{ceofdm} preserves a unit-magnitude sample envelope, but its Hermitian-symmetric structure makes dedicated sensing pilots costly, and additive pilots disrupt this envelope. We show that injecting a known chirp, here a \gls{zc} sequence, into the CE-OFDM phase enables payload-blind bistatic delay--Doppler sensing without dedicated sensing frames or an explicit pilot power split: the residual carrier, given by the ensemble mean under the adopted payload model, becomes an implicit superimposed pilot for a receive-only node. A Doppler-aligned receiver integrates this reference coherently across frames while the payload residual averages down. In contrast, \gls{pit} frame averaging, designed for quasi-static channel estimation, loses about 32~dB for the simulated target with 16 averaged frames. Closed-form approximations characterize the payload-induced floor, known-cell detection probability and direct-path tolerance. In the noise-limited simulations, the receiver incurs a 2.8~dB detection gap relative to a payload-aware matched filter without interference cancellation and matches a dedicated-pilot benchmark of equal reference energy without its 53\% frame overhead.
\end{abstract}


\section{Introduction}

\Gls{ceofdm} is a waveform with potential for application in transmitters constrained by power-amplifier nonlinearities and efficiency, since it uses a unit-magnitude envelope which keeps the amplifier operating near saturation without the required power back-off on conventional \gls{ofdm}~\cite{thompson2008}.
However, the real-valued phase signal requires Hermitian symmetry, leaving fewer than half of the subcarriers for independent data symbols.

This reduced rate becomes even more relevant in bistatic sensing scenarios, where a spatially separated receiver needs a known transmit reference to perform matched filtering~\cite{liu2022isac,brunner2025bistatic,keskin2025}.
The usual approach of inserting dedicated pilot subcarriers or training frames provides a clean reference, but reduces the throughput of a system.
Data-aided systems avoid pilot overhead by decoding the payload to reconstruct the reference~\cite{keskin2025}. However, they require a decodable direct path and a full communication demodulator at the sensing receiver.

\Gls{st} avoids both costs and has recently been shown to enable payload-blind delay--Doppler sensing in \gls{ofdm}: M\'endez-Monsanto \emph{et al.}~\cite{mendez2025superimposed} add a \gls{zc} pilot~\cite{chu1972} to the \gls{ofdm} signal and detect targets with a Doppler-matched correlator bank, at the cost of an explicit pilot power fraction.
Kumar \emph{et al.}~\cite{kumar2026} superimpose a chirp on \gls{ofdm} with an explicit power split and use a chirp-only matched filter for sensing, including bistatically; their waveform reduces the \gls{papr} but does not have a constant envelope.
Additive \gls{st}, however, destroys the constant envelope of \gls{ceofdm}~\cite{chenhu2023}.
\Gls{pit}~\cite{chenhu2023} instead injects a deterministic sequence into the \gls{ceofdm} phase: averaging received frames removes the payload and leaves the sequence scaled by the residual carrier of \gls{ceofdm}, from which the channel is estimated.
\gls{pit} was designed for quasi-static channels, and even its frequency-modulated \gls{ofdm} variant, intended for high mobility, assumes a channel that is constant over the averaged slot~\cite{chenhu2023spawc}.
With a moving target, however, the reference rotates from frame to frame, so this average attenuates the reference and can even cancel it; the frames must therefore first be aligned in Doppler.

In the radar literature, Zhang \emph{et al.}~\cite{zhang2019ceofdmpm} add a scaled data phase to a pseudo-random radar phase in a monostatic radar that knows its transmitted waveform.
Frame-to-frame payload changes then cause range-sidelobe modulation, which they reduce by shrinking the data phase at the expense of the bit error rate. %; we instead average this payload residual out.
\Gls{ceofdm} radar has also been combined with a \gls{lfm} chirp~\cite{bai2022}, data-aided processing~\cite{huang2018radcom} or chirp subcarriers~\cite{dida2016}, and frequency-modulated \gls{ofdm} has been analyzed for monostatic sensing with a data-aided range receiver~\cite{bouziane2025,bouziane2026} %, leaving bistatic operation to future work~\cite{bouziane2026}.
To our knowledge, none of these works analyzes the detection performance of payload-blind bistatic sensing with a constant-envelope waveform.

In this work, we extend \gls{pit}, originally developed for channel estimation to sensing, by recovering the waveform implicit sensing reference through a Doppler-aligned averaging. While Doppler alignment itself is a well-established radar technique, our contribution lies in adapting the \gls{pit} essence to bistatic sensing and analytically characterizing the resulting payload-induced interference floor, detection probability, and direct-path tolerance.

Unlike the explicit pilots of~\cite{mendez2025superimposed,kumar2026}, the residual carrier costs the payload neither power nor bit-error-rate performance at a given modulation index, and it preserves the constant envelope of the transmitted samples.

Our contributions can be summarized as follows:
\begin{itemize}
\item We show that injecting a full-length chirp reference, here a \gls{zc} sequence, into the \gls{ceofdm} phase turns the residual carrier into an implicit superimposed pilot for payload-blind bistatic delay--Doppler sensing.
\item We derive closed-form approximations for the payload-induced floor, \gls{pfr} and known-cell detection probability, and compare the receiver with a payload-aware matched filter without interference cancellation and energy-matched dedicated pilots.
\item We quantify the direct-path tolerance of payload-blind bistatic sensing under the adopted interference model.
\end{itemize}

The remainder of this work is organized as follows: Section \ref{sec:model} describes the considered system model, while Section \ref{sec:receiver} presents the proposed receiver scheme together with a performance analysis. Numerical results are given in Section \ref{sec:results}; finally, Section V contains the concluding remarks.

\noindent\emph{Notation:} $\mathbb{E}\{\cdot\}$, $\Re\{\cdot\}$ and $(\cdot)^*$ denote expectation, real part and complex conjugation; $(\cdot)_N$ denotes modulo-$N$ indexing; and $\mathcal{N}(0,1)$ and $\mathcal{CN}(0,\sigma^2)$ denote the standard real and circularly symmetric complex Gaussian distributions.

\section{Signal Model}
\label{sec:model}

\begin{figure*}[!t]
\centering
% Figure 1 source. Included inside the manuscript's figure* environment.
% Requires TikZ and the arrows.meta library from the manuscript preamble.
\begin{tikzpicture}[x=1cm,y=1cm,font=\footnotesize,
 block/.style={draw,rounded corners=1.5pt,align=center,anchor=west,text width=2.25cm,
 minimum height=0.92cm,inner sep=3pt},
 flow/.style={-{Latex[length=1.7mm]},semithick},
 heading/.style={anchor=west,font=\footnotesize\bfseries}]
\node[heading] at (0,0.78) {Transmitter};
% Position blocks by their left edges so unequal widths cannot misalign rows.
\node[block] (qam) at (0,0) {Payload bits\\QAM mapping};
\node[block] (idft) at (3,0) {Hermitian mapping\\IDFT};
\node[block] (ce) at (6,0) {CE phase mapping\\$e^{j\gamma u_b[n]}$};
\node[block] (chirp) at (9,0) {Chirp phase injection\\$a[n]s_b[n]$};
\node[block] (cp) at (12,0) {Cyclic prefix\\insertion};
\draw[flow] (qam)--(idft);
\draw[flow] (idft)--(ce);
\draw[flow] (ce)--(chirp);
\draw[flow] (chirp)--(cp);
% A shared channel bar at the right feeds both right-to-left receivers.
\node[draw,rounded corners=1.5pt,minimum width=0.9cm,
 minimum height=4.85cm,inner sep=0pt] (channel) at (15.3,-1.9) {};
\node[rotate=90] at (channel.center) {Doubly dispersive channel + noise};
\draw[flow] (cp.east)--(channel.west |- cp.east);
\node[heading] at (0,-1.12) {Payload-blind sensing receiver};
\node[block] (front) at (12,-1.9) {CP removal\\received $r_b[n]$};
\node[block] (blind) at (8,-1.9) {Doppler-compensated\\chirp correlation};
\node[block] (avg) at (4,-1.9) {Doppler alignment\\coherent average};
\node[block] (peak) at (0,-1.9) {Delay--Doppler\\peak search};
\draw[flow] (channel.west |- front.east)--(front.east);
\draw[flow] (front)--(blind);
\draw[flow] (blind)--(avg);
\draw[flow] (avg)--(peak);
\node[heading] at (0,-3.02) {Communication receiver};
% The narrower communication blocks share the same outer row boundaries.
\node[block,text width=1.95cm] (commfront) at (12.30,-3.8) {CP removal};
\node[block,text width=1.95cm] (equalizer) at (9.84,-3.8) {Channel\\equalization};
\node[block,text width=1.95cm] (dechirp) at (7.38,-3.8) {Chirp removal\\multiply by $a^*[n]$};
\node[block,text width=1.95cm] (phase) at (4.92,-3.8) {Phase unwrapping};
\node[block,text width=1.95cm] (dft) at (2.46,-3.8) {DFT};
\node[block,text width=1.95cm] (bits) at (0,-3.8) {QAM decisions\\bit demapping};
\draw[flow] (channel.west |- commfront.east)--(commfront.east);
\draw[flow] (commfront)--(equalizer);
\draw[flow] (equalizer)--(dechirp);
\draw[flow] (dechirp)--(phase);
\draw[flow] (phase)--(dft);
\draw[flow] (dft)--(bits);
\end{tikzpicture}
\caption{Transmitter (top), payload-blind sensing receiver (middle) and communication receiver (bottom). For each candidate Doppler $f$, the sensing receiver correlates each Doppler-compensated frame $r_b[n]e^{-j2\pi fn/F_s}$ with the injected chirp and then averages the Doppler-aligned correlations, which is equivalent to~\eqref{eq:doppler_pit_average}--\eqref{eq:blind} by linearity.}
\label{fig:system}
\end{figure*}
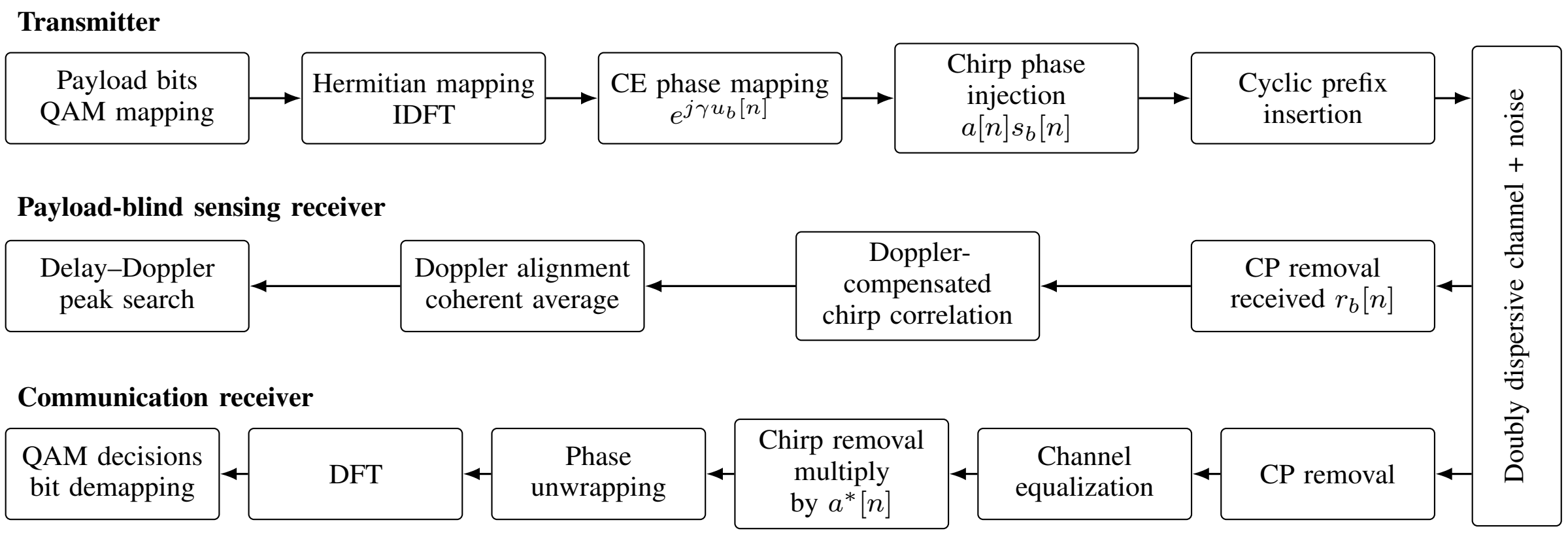

\subsection{Chirp-Injected \gls{ceofdm} Transmitter}

The transmitter embeds a known reference in the phase of \gls{ceofdm} without reserving subcarriers (Fig.~\ref{fig:system}, top).
Consider $B$ consecutive frames $b\in\{0,\dots,B-1\}$, each carrying $N_d$ independent unit-power $M$-ary \gls{qam} symbols $\{d_{b,k}\}_{k=1}^{N_d}$.
The symbols are mapped to the positive-frequency bins of an $N$-point inverse \gls{dft} with $N_d\leq N/2-1$, their conjugates to the negative-frequency bins, and the \gls{dc} and Nyquist bins are set to zero, which yields the real precursor~\cite{thompson2008}
\begin{equation}
u_b[n] = \sqrt{\frac{2}{N_d}}\sum_{k=1}^{N_d}\Re\left\{d_{b,k}e^{j2\pi kn/N}\right\}, \quad 0 \leq n < N,
\label{eq:precursor}
\end{equation}
normalized so that $\mathbb{E}\{u_b^2[n]\}=1$.
\gls{ceofdm} phase modulation~\cite{thompson2008} yields $s_b[n]=e^{j\gamma u_b[n]}$, where the modulation index $\gamma$ sets the data-bearing phase excursion.

Following the \gls{pit} principle~\cite{chenhu2023}, we inject a frame-invariant phase sequence into $s_b[n]$.
The injected reference can be any unit-modulus sequence $a[n]=e^{j\varphi[n]}$ with good periodic autocorrelation; here, the centered even-length \gls{zc} sequence is used, a discrete chirp~\cite{chu1972}:
\begin{equation}
a[n] = e^{j\pi q (n-N/2)^2/N}, \quad 0 \leq n < N,
\label{eq:chirp}
\end{equation}
whose shift by $N/2$ centers the frequency sweep within the band. For an integer root $q$ coprime with $N$, $a[n]$ has unit magnitude and ideal periodic autocorrelation~\cite{chu1972}:
\begin{equation}
R_a[\ell] \triangleq \frac{1}{N}\sum_{n=0}^{N-1} a[n]a^*[(n-\ell)_N] =
\begin{cases}
1, & \ell = 0,\\
0, & 1 \leq \ell < N.
\end{cases}
\label{eq:zc_autocorrelation}
\end{equation}

The \gls{pit} scheme~\cite{chenhu2023} repeats a short phase sequence several times per frame for channel estimation; we instead inject a single length-$N$ sequence, so that the delay response is unambiguous over the whole frame, and use $q=1$, since $q$ only sets the delay--Doppler coupling of the \gls{zc} sequence~\cite{mendez2025superimposed}, which the intra-frame compensation in~\eqref{eq:blind} removes at the target Doppler.
The transmitted frame is
\begin{equation}
x_b[n] = a[n]s_b[n] = e^{j(\varphi[n] + \gamma u_b[n])}.
\label{eq:transmitted}
\end{equation}

Since $|x_b[n]|=1$ for all $n$,
\begin{equation}
\mathrm{PAPR}\{x_b\}=\frac{\max_n|x_b[n]|^2}{\frac{1}{N}\sum_{n=0}^{N-1}|x_b[n]|^2}=1\;(0\text{ dB}),
\end{equation}
so the injected reference preserves the constant envelope of the transmitted samples, as in~\cite{chenhu2023}. As for any \gls{ceofdm} signal sampled at the Nyquist rate, the interpolated continuous-time envelope is only approximately constant; with oversampling, a band-matched reference, such as \gls{lfm} or \gls{nlfm} chirps swept over the data band, could keep it nearly constant at the cost of nonzero range sidelobes.

At the communication receiver (Fig.~\ref{fig:system}, bottom), multiplying by $a^*[n]$ after equalization removes the known chirp and restores the \gls{ceofdm} signal exactly, so with ideal channel knowledge the bit error rate is that of \gls{ceofdm}~\cite{thompson2008}.

\subsection{Reference and Payload Residual}

The proposed first-order coherent receiver correlates against the payload-independent mean of $x_b[n]$.
For large $N_d$, the central limit theorem justifies modeling $u_b[n]\sim\mathcal{N}(0,1)$~\cite{thompson2008,chenhu2023}, and the Gaussian characteristic function gives~\cite[Eq.~(31)]{chenhu2023}
\begin{equation}
\mathbb{E}\{s_b[n]\} = \mathbb{E}\left\{e^{j\gamma u_b[n]}\right\} \simeq e^{-\gamma^2/2} \triangleq \mu_\gamma.
\label{eq:mean}
\end{equation}

Since $a[n]$ is deterministic, each frame decomposes into a scaled chirp and a payload residual that is zero-mean to the accuracy of~\eqref{eq:mean},
\begin{equation}
x_b[n] = \mu_\gamma a[n] + v_b[n], \qquad v_b[n] \triangleq a[n](s_b[n]-\mu_\gamma),
\label{eq:referenced}
\end{equation}
with $\mathbb{E}\{|v_b[n]|^2\} = 1-\mu_\gamma^2$.

The residual carrier $\mu_\gamma a[n]$ thus holds a fraction $\mu_\gamma^2=e^{-\gamma^2}$ of the power ($52.7\%$ for $\gamma=0.8$); in plain \gls{ceofdm} this component is an unmodulated tone at \gls{dc}, which the injection spreads over the band.
Averaging $B$ independent frames~\cite[Eq.~(34)]{chenhu2023} we obtain
\begin{equation}
\bar{x}[n] \triangleq \frac{1}{B}\sum_{b=0}^{B-1} x_b[n] = \mu_\gamma a[n] + \frac{1}{B}\sum_{b=0}^{B-1} v_b[n],
\label{eq:decaying_vb}
\end{equation}
which preserves the chirp while the residual variance decays as $(1-\mu_\gamma^2)/B$.

\section{Channel, Receiver and Performance Analysis}
\label{sec:receiver}

\subsection{Doubly Dispersive Channel and \gls{pit} Averaging}

We consider a single bistatic link between a \gls{ceofdm} transmitter and a receive-only sensing node that knows $N$, $q$ and the frame timing, but not the payload.
A \gls{cp} of length $N_{\text{cp}}$ is prepended to each frame, $\tilde{x}_b[n] = x_b[(n)_N]$ for $-N_{\text{cp}} \leq n < N$, and the frame propagates through a doubly dispersive channel with a direct path $p=0$ and $P$ target paths,
\begin{equation}
\tilde{r}_b[n] = \sum_{p=0}^{P} \alpha_p e^{j2\pi f_p(n/F_s + bT_f)} \tilde{x}_b[n - \tau_p] + w_b[n],
\label{eq:linear_channel}
\end{equation}
where $\alpha_p$, $\tau_p$ and $f_p$ are the complex gain, integer delay and Doppler of path $p$, $F_s$ is the sampling rate, $T_f = (N + N_{\text{cp}})/F_s$ is the frame duration, and $w_b[n] \sim \mathcal{CN}(0, \sigma_w^2)$ is i.i.d.\ over $n$ and $b$.

When $N_{\text{cp}} \geq \max_p \tau_p$, removing the \gls{cp} yields the circular model
\begin{equation}
r_b[n] = \sum_{p=0}^{P} \alpha_p e^{j2\pi f_p(n/F_s + bT_f)} x_b[(n-\tau_p)_N] + w_b[n].
\label{eq:channel}
\end{equation}

The node is frame-synchronized and compensates the carrier offset using the direct path, so that $\tau_0=f_0=0$ and $\tau_p$ and $f_p$ are the bistatic delay and Doppler of target $p\geq1$ relative to the direct path; a residual carrier offset appears as a common Doppler shift and is absorbed by the Doppler search.
We further assume that Doppler shifts lie on the search grid, that $\alpha_p$ and $f_p$ remain constant over the coherent interval $BT_f$, and that scrambled payloads make $u_b[n]$ independent across frames.

The average~\eqref{eq:decaying_vb} assumes a reference that is identical in every frame.
Applied to sensing, the \gls{pit} average~\cite{chenhu2023}, $\bar{r}_{\text{PIT}}[n] = (1/B)\sum_{b} r_b[n]$, therefore quantifies the Doppler sensitivity of injected training: the reference of target $p$ in frame $b$ is rotated by $e^{j2\pi f_p b T_f}$, so substituting~\eqref{eq:referenced} and~\eqref{eq:channel} scales it by the Dirichlet kernel
\begin{equation}
D_B(f_p T_f) \triangleq \frac{1}{B}\sum_{b=0}^{B-1} e^{j2\pi f_p b T_f}= e^{j\pi f_p(B-1)T_f}\frac{\sin(\pi f_p B T_f)}{B\sin(\pi f_p T_f)},
\label{eq:pit_loss}
\end{equation}
whose magnitude decays as $1/B$ and vanishes whenever $f_p B T_f$ is a nonzero integer, attenuating or canceling moving targets.

\subsection{Doppler-Aligned Payload-Blind Receiver}

For each candidate Doppler $f$, the proposed receiver (Fig.~\ref{fig:system}, middle) first aligns the frames in slow time and averages them, resulting in
\begin{equation}
\bar{r}_f[n] = \frac{1}{B}\sum_{b=0}^{B-1} r_b[n]e^{-j2\pi f b T_f},
\label{eq:doppler_pit_average}
\end{equation}
which reduces to the \gls{pit} average for $f=0$ and gives $D_B(0)=1$ for $f = f_p$.

\begin{subequations}
\label{eq:blind_both}
For each candidate delay $\tau$, $\bar{r}_f[n]$ is then correlated in fast time with the reference $a[n]$ while compensating the intra-frame Doppler:
\begin{eqnarray}
Z_B(\tau, f) &&\hspace{-4ex}= \frac{1}{N}\sum_{n=0}^{N-1} \bar{r}_f[n]\,a^*[(n\!-\!\tau)_N]\,e^{-j2\pi f n/F_s}
\label{eq:blind}\\
&&\hspace{-12ex}= \frac{1}{BN}\sum_{b=0}^{B-1}\sum_{n=0}^{N-1} r_b[n]\,a^*[(n\!-\!\tau)_N]\!\times\! e^{-j2\pi f\left(\tfrac{n}{F_s}+bT_f\right)}\!\!.\qquad
\label{eq:blind_combined}
\end{eqnarray}
\end{subequations}

This is the classical fast-time/slow-time processing of pulse-Doppler and \gls{ofdm} radar~\cite{sturm2011} and the Doppler-matched correlator of additive \gls{st}~\cite{mendez2025superimposed}, applied here to the residual carrier, so no payload knowledge is needed.

For the uncompensated baselines, define the per-frame correlation
\begin{equation}
c_b(\tau)=\frac{1}{N}\sum_{n=0}^{N-1}r_b[n]a^*[(n-\tau)_N].
\label{eq:frame_corr}
\end{equation}
The non-coherent baseline averages powers,
$T_{\mathrm{NC}}(\tau)=B^{-1}\sum_b|c_b(\tau)|^2$,
whereas \gls{pit} averaging uses
$T_{\mathrm{PIT}}(\tau)=|B^{-1}\sum_b c_b(\tau)|^2$.
Both avoid a Doppler search; \gls{pit} averaging retains phase and can cancel moving targets.

\subsection{Performance Analysis}

Taking the expectation of~\eqref{eq:blind_combined} over the payload, using~\eqref{eq:referenced} and~\eqref{eq:channel}, gives, with $\nu_p\triangleq f_p-f$ and up to a constant phase per path,
\begin{equation}
\mathbb{E}\{Z_B(\tau,f)\}\simeq\mu_\gamma{\textstyle\sum_{p}}\alpha_p\,\chi_a(\tau-\tau_p,\nu_p/F_s)\,D_B(\nu_pT_f),
\label{eq:af}
\end{equation}
where $\chi_a(\ell,\theta)\triangleq(1/N)\sum_{n}a[n]a^*[(n-\ell)_N]e^{j2\pi\theta n}$ is the periodic ambiguity function of the reference.

Its zero-Doppler cut is $R_a[\ell]$ in~\eqref{eq:zc_autocorrelation}, and its Doppler cut at the true delay is $D_N(\cdot)D_B(\cdot)$, with $D_N(\theta)\triangleq(1/N)\sum_n e^{j2\pi\theta n}$. The first nulls of $D_B$, at $\pm1/(BT_f)${\color{blue},} set the Doppler resolution.
At $(\tau_p,f_p)$, $|\mathbb{E}\{Z_B\}|\simeq|\alpha_p|\mu_\gamma$.

Away from the path delays, we treat the payload residual as white (which holds without oversampling, since $N_d\approx N/2$), neglect the cross-terms between the residuals of different paths, and use the independence of payloads and noise across frames.
The interference-plus-noise power of~\eqref{eq:blind_combined} is then
\begin{equation}
\sigma_I^2 \simeq \frac{(1-\mu_\gamma^2)\sum_{p}|\alpha_p|^2+\sigma_w^2}{NB},
\label{eq:floor}
\end{equation}
which has the $1/(NB)$ form of the additive-\gls{st} interference in~\cite{mendez2025superimposed}, with the payload fraction fixed at $1-\mu_\gamma^2$ by the modulation index instead of chosen by a power split.

We call the receiver \emph{noise-limited} when $\sigma_w^2\gg(1-\mu_\gamma^2)\sum_p|\alpha_p|^2$, the usual low \gls{snr} operating point of sensing, and \emph{payload-limited} in the opposite case, which occurs for strong echoes or, in bistatic sensing, a strong direct path.
In the payload-limited regime, the \gls{pfr} no longer increases with the \gls{snr}, and only longer integration improves detection at a given $\gamma$.
Equations~\eqref{eq:floor}--\eqref{eq:dtr} below use only $|a[n]|=1$ and a white residual, so they hold for any injected reference up to the deterministic sidelobes of its ambiguity function $\chi_a$ in~\eqref{eq:af}, which add to the floor near strong paths and, for the \gls{zc} sequence, vanish only at zero Doppler offset.

We measure sensing performance by the \gls{pfr} of path $p$ at candidate Doppler $f$:
\begin{equation}
\Gamma_p(f)=\frac{\mu_\gamma^2|\alpha_p|^2\,|D_N(\nu_p/F_s)|^2\,|D_B(\nu_pT_f)|^2}{\sigma_I^2}.
\label{eq:pfr}
\end{equation}

%The proposed receiver ($f=f_p$) attains $NB\mu_\gamma^2|\alpha_p|^2/[(1-\mu_\gamma^2)\sum_p|\alpha_p|^2+\sigma_w^2]$, i.e., a $10\log_{10}B$~dB gain, because the reference adds coherently over the $NB$ samples while the payload residual and noise add incoherently.
%\gls{pit} averaging ($f=0$) is further scaled by $|D_N(f_p/F_s)D_B(f_pT_f)|^2$.
%A payload-aware genie that correlates against $x_b[n]$, the upper bound of data-aided sensing~\cite{keskin2025}, attains~\eqref{eq:pfr} with $\mu_\gamma^2\to1$ and $1-\mu_\gamma^2\to1-\mu_\gamma^4$, so the price of blindness is $10\log_{10}e^{\gamma^2}$ when noise-limited and $10\log_{10}[1/(\mu_\gamma^2(1+\mu_\gamma^2))]$ when payload-limited. \henrique{Too much math embeeded in the text?}
At $f=f_p$, the reference adds coherently while the independent payload residual and noise average down, yielding the $10\log_{10}B$~dB integration gain in~\eqref{eq:pfr}.
\gls{pit} averaging ($f=0$) is further scaled by $|D_N(f_p/F_s)D_B(f_pT_f)|^2$.
A payload-aware matched filter without interference cancellation (the genie) correlates against $x_b[n]$. Its \gls{pfr} is approximated by~\eqref{eq:pfr} with $\mu_\gamma^2\to1$ and $1-\mu_\gamma^2\to1-\mu_\gamma^4$. For $\gamma=0.8$, the resulting \gls{pfr} gap is $2.8$~dB when noise-limited and approaches $0.9$~dB when payload-limited.

For comparison, we also introduce energy-matched pilots, where $K=\mathrm{round}(\mu_\gamma^2B)$ of the $B$ frames carry only the reference ($\gamma=0$) and are processed by~\eqref{eq:doppler_pit_average}--\nolinebreak\eqref{eq:blind} with $B$ replaced by $K$, so that they convey the same reference energy as the proposed scheme at the cost of a fraction $K/B\approx\mu_\gamma^2$ of the frames.
Free of payload residual, they attain $\Gamma_p=NK|\alpha_p|^2/\sigma_w^2$, which equals the proposed \gls{pfr} when noise-limited and exceeds it when payload-limited.

With the interference approximated as circularly symmetric complex Gaussian, a square-law detector with per-cell false-alarm probability $P_{fa}$ approximately attains~\cite{richards2022}
\begin{equation}
P_d=Q_1\big(\sqrt{2\Gamma_p},\sqrt{-2\ln P_{fa}}\big),
\label{eq:pd}
\end{equation}
where $Q_1$ is the Marcum $Q$-function.
Equation~\eqref{eq:floor} describes the off-target cells that set the detection threshold; their simulated power is exponentially distributed in both regimes, as assumed.
At a target's own cell, the part of its residual that is linear in $u_b$ ($\gamma^2\mu_\gamma^2$ of the power $1-\mu_\gamma^2$) vanishes because $u_b$ has no \gls{dc} component, so that residual is about $5.4$~dB smaller for $\gamma=0.8$ and~\eqref{eq:pd} is then conservative.

Since the direct path enters~\eqref{eq:floor} through $\sum_p|\alpha_p|^2$, a target attains the required \gls{pfr} $\Gamma_{\mathrm{req}}$, neglecting noise and the other targets, only if the \gls{dtr} $\mathrm{DTR}=|\alpha_0|^2/|\alpha_p|^2$ satisfies
\begin{equation}
\mathrm{DTR}\lesssim\mathrm{DTR}_{\max}=\frac{NB\mu_\gamma^2}{(1-\mu_\gamma^2)\,\Gamma_{\mathrm{req}}}.
\label{eq:dtr}
\end{equation}

For the genie, the same substitutions raise $\mathrm{DTR}_{\max}$ by $0.9$~dB.

\section{Numerical Results}
\label{sec:results}

\begin{table}[!t]
\caption{Simulation Parameters}
\label{tab:simul_params}
\centering
\begin{tabular}{@{}ll@{}}
\toprule
Parameter & Value\\
\midrule
\gls{dft} size $N$; payload symbols $N_d$ & $1024$; $511$ (no oversampling)\\
Modulation; modulation index $\gamma$ & 16-\gls{qam}; $0.8$\\
Sampling rate $F_s$; \gls{cp} length $N_{\text{cp}}$ & $4$~MHz; $256$ samples\\
Frame duration $T_f$; \gls{zc} root $q$ & $320$~\textmu s; $1$\\
Target delays $\tau_1,\tau_2,\tau_3$ ($P=3$; samples) & $23$, $44$, $50$\\
Target Dopplers $f_1,f_2,f_3$ & $+600$, $-450$, $+900$~Hz\\
Target gains $|\alpha_1|^2,|\alpha_2|^2,|\alpha_3|^2$ & $0$, $-4$, $-8$~dB\\
Per-cell $P_{fa}$; Monte Carlo trials & $10^{-4}$; $200$--$400$ per point\\
\bottomrule
\end{tabular}
\end{table}

We validate the analysis with the parameters of Table~\ref{tab:simul_params}; the per-sample \gls{snr} of target~1 is $S_1=|\alpha_1|^2/\sigma_w^2$, and the direct path is absent ($\alpha_0=0$, ideal synchronization) except in Fig.~\ref{fig:pd}(b).
Theory curves use the exact mean of the finite 16-\gls{qam} constellation (within $0.001$~dB of $e^{-\gamma^2/2}$), and Monte Carlo markers carry $95\%$ confidence intervals.

\subsection{Delay and Doppler Response}

\begin{figure}[!t]
\centering
% Fig. 2: mean correlation power, normalized to the genie peak. Data: data/cuts_*.dat (export_tikz_data_v3.m).
\pgfplotstableread{figures_v3/data/cuts_consts.dat}\cutsconsts
\pgfplotstablegetelem{0}{blindfl1}\of\cutsconsts\let\blfa\pgfplotsretval
\pgfplotstablegetelem{0}{blindfl16}\of\cutsconsts\let\blfb\pgfplotsretval
\pgfplotstablegetelem{0}{geniefl1}\of\cutsconsts\let\gefa\pgfplotsretval
\pgfplotstablegetelem{0}{geniefl16}\of\cutsconsts\let\gefb\pgfplotsretval
\pgfplotstablegetelem{0}{pity}\of\cutsconsts\let\pity\pgfplotsretval
\begin{tikzpicture}
\begin{groupplot}[group style={group size=1 by 2, vertical sep=27pt}, v3axis, height=79pt]
% (a) delay cut at zero Doppler
\nextgroupplot[xmin=-60, xmax=60, ymin=-50, ymax=2, xtick={-60,-40,...,60}, ytick={-50,-40,...,0},
  xlabel={Delay $\tau$ (samples)}, ylabel={Power (dB)},
  legend style={at={(0.01,0.98)}, anchor=north west}]
\addplot[color=cgenie, line width=0.8pt] table[x=tau, y=genie1] {figures_v3/data/cuts_delay.dat};
\addlegendentry{Genie, $B=1$}
\addplot[color=cblind, line width=0.8pt] table[x=tau, y=blind1] {figures_v3/data/cuts_delay.dat};
\addlegendentry{Proposed, $B=1$}
\addplot[color=cgenie, line width=0.8pt, dashed] table[x=tau, y=genie16] {figures_v3/data/cuts_delay.dat};
\addlegendentry{Genie, $B=16$}
\addplot[color=cblind, line width=0.8pt, dashed] table[x=tau, y=blind16] {figures_v3/data/cuts_delay.dat};
\addlegendentry{Proposed, $B=16$}
\addplot[color=ctheory, line width=0.7pt, densely dotted, domain=-60:60, samples=2] {\blfa};
\addlegendentry{Theory}
\addplot[color=ctheory, line width=0.7pt, densely dotted, domain=-60:60, samples=2, forget plot] {\blfb};
\addplot[color=ctheory, line width=0.7pt, densely dotted, domain=-60:60, samples=2, forget plot] {\gefa};
\addplot[color=ctheory, line width=0.7pt, densely dotted, domain=-60:60, samples=2, forget plot] {\gefb};
\node[v3panel, anchor=north east] at (rel axis cs:0.99,0.98) {(a)};
% (b) Doppler cut at the delay of the 600 Hz target
\nextgroupplot[xmin=-1.5625, xmax=1.5625, ymin=-50, ymax=8, xtick={-1.5,-1,...,1.5}, ytick={-50,-40,...,0},
  xticklabel style={/pgf/number format/.cd, fixed, precision=1},
  xlabel={Doppler $f$ (kHz)}, ylabel={Power (dB)}]
\addplot[color=cgenie, line width=0.8pt] table[x=fk, y=genie1] {figures_v3/data/cuts_doppler.dat};
\addplot[color=cblind, line width=0.8pt] table[x=fk, y=blind1] {figures_v3/data/cuts_doppler.dat};
\addplot[color=cgenie, line width=0.8pt, dashed] table[x=fk, y=genie16] {figures_v3/data/cuts_doppler.dat};
\addplot[color=cblind, line width=0.8pt, dashed] table[x=fk, y=blind16] {figures_v3/data/cuts_doppler.dat};
\addplot[color=ctheory, line width=0.7pt, densely dotted] table[x=fk, y=th1] {figures_v3/data/cuts_doppler.dat};
\addplot[color=ctheory, line width=0.7pt, densely dotted] table[x=fk, y=th16] {figures_v3/data/cuts_doppler.dat};
\draw[black, line width=0.4pt] (axis cs:-0.9,-21.5) -- (axis cs:-0.03,\pity+0.8);
\node[font=\scriptsize, anchor=south west, inner sep=1pt] at (axis cs:-1.47,-21.5) {PIT ($f=0$)};
\addplot[only marks, mark=*, mark size=1.6pt, mark options={fill=white, draw=black, line width=0.6pt}]
  coordinates {(0,\pity)};
\node[v3panel, anchor=north west] at (rel axis cs:0.01,0.98) {(b)};
\end{groupplot}
\end{tikzpicture}
\vspace{-4ex}
\caption{Mean correlation power without noise for a single target, normalized to the genie peak, for $B=1$ and $16$. (a) Delay cut at zero Doppler for a static target (mean over $10^4$ realizations). (b) Doppler cut at the delay of a $600$~Hz target; the marker shows where \gls{pit} averages ($f=0$, $B=16$). Dotted lines: theory from~\eqref{eq:af} and~\eqref{eq:floor}.}
\label{fig:cuts}
\end{figure}
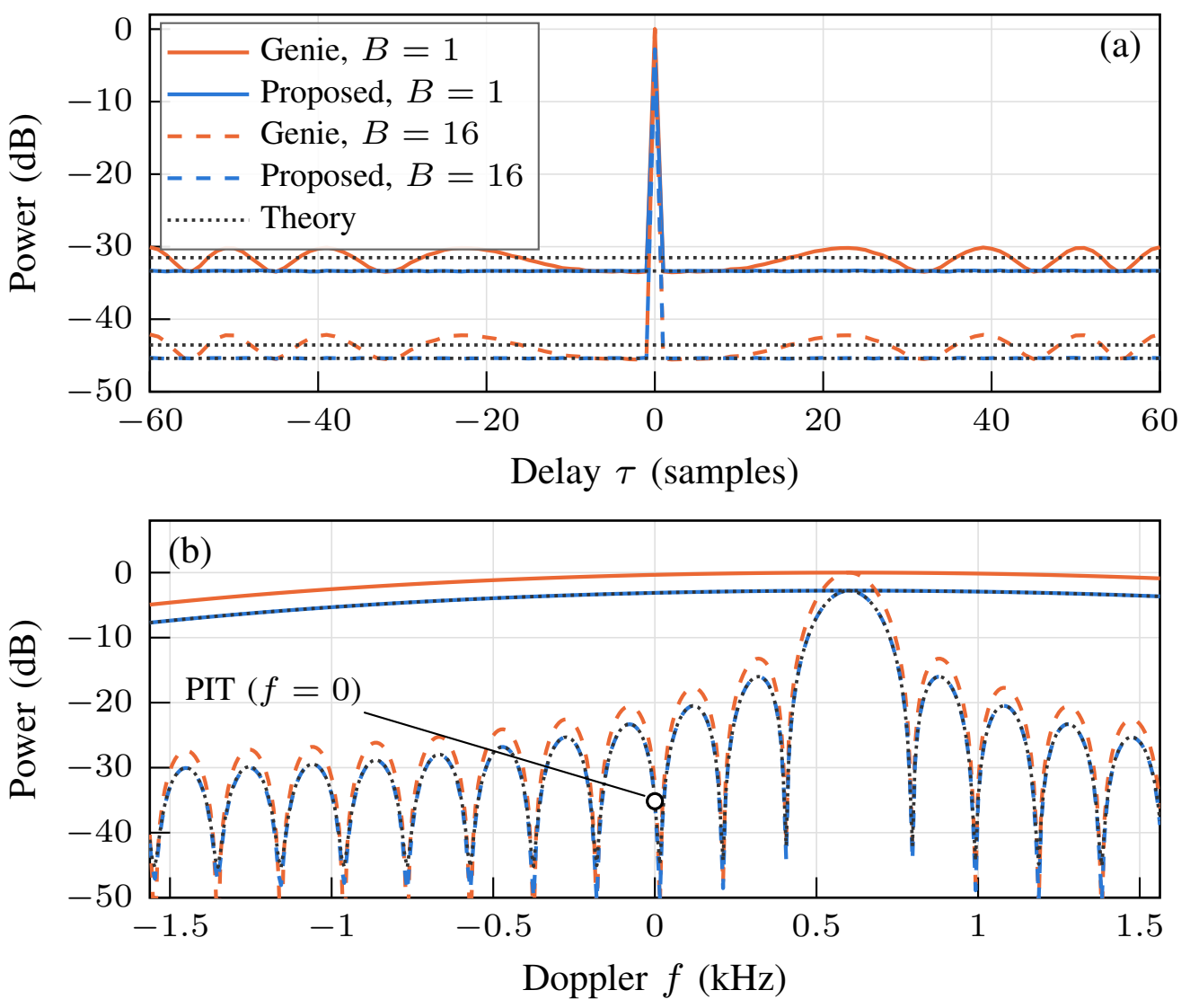

Fig.~\ref{fig:cuts} compares the proposed correlator~\eqref{eq:blind_combined} with the genie of Section~\ref{sec:receiver}-C.
In the delay cut, both share the \gls{zc} mainlobe, and the proposed peak lies $\mu_\gamma^2$ ($2.8$~dB) below the genie peak.
Both floors are set by the payload and fall by $10\log_{10}B$ ($12$~dB at $B=16$).
For $B=1$, the proposed floor lies $30.6$~dB below its own peak, as predicted by~\eqref{eq:floor}, and the genie floor lies, on average over the delay, only $0.9$~dB further below the genie peak because of its own payload self-clutter.
The Doppler cut follows~\eqref{eq:af}: its first nulls move to $\pm1/(BT_f)=\pm195$~Hz around the peak at $B=16$, and the response at $f=0$, where \gls{pit} averages, lies $32$~dB below the peak.
For $B=1$, the cut is nearly flat (one frame resolves only $F_s/N=3.9$~kHz), so Doppler selectivity and \gls{pit}'s loss arise across frames.

\subsection{Doppler Alignment versus \gls{pit} Averaging}

\begin{figure}[!t]
\centering
% Fig. 3: PFR of target 1 versus B. Lines: theory (data/pfr_th_k.dat); markers: Monte Carlo with 95% intervals
% (data/pfr_mc_k.dat). Data written by matlab_v3/export_tikz_data_v3.m.
\begin{tikzpicture}
\begin{groupplot}[group style={group size=1 by 2, vertical sep=14pt}, v3axis, height=74pt,
  xmode=log, xmin=1, xmax=64, xtick={1,2,4,8,16,32,64}, xticklabels={1,2,4,8,16,32,64}, minor xtick={},
  ylabel={PFR (dB)}]
\nextgroupplot[ymin=-30, ymax=30, ytick={-20,0,20},
  legend columns=2, legend style={at={(0.5,1.03)}, anchor=south, /tikz/every even column/.append style={column sep=6pt}}]
\addlegendimage{v3leg=mgenie}\addlegendentry{Genie (payload known)}
\addlegendimage{v3leg=mpilot}\addlegendentry{Energy-matched pilots}
\addlegendimage{v3leg=mblind}\addlegendentry{Proposed (blind)}
\addlegendimage{v3leg=mpit}\addlegendentry{PIT averaging ($f=0$)}
\pgfplotsinvokeforeach{genie,pilot,blind,pit}{
  \addplot[v3th=m#1] table[x=B, y=#1] {figures_v3/data/pfr_th_1.dat};
  \addplot[v3mc=m#1] table[x=B, y=#1, y error minus=#1m, y error plus=#1p] {figures_v3/data/pfr_mc_1.dat};
}
\node[v3panel, anchor=north west] at (rel axis cs:0.01,0.98) {(a) $S_1=-25$~dB};
\nextgroupplot[ymin=-5, ymax=60, ytick={0,20,40,60}, xlabel={Number of frames $B$}]
\pgfplotsinvokeforeach{genie,pilot,blind,pit}{
  \addplot[v3th=m#1] table[x=B, y=#1] {figures_v3/data/pfr_th_2.dat};
  \addplot[v3mc=m#1] table[x=B, y=#1, y error minus=#1m, y error plus=#1p] {figures_v3/data/pfr_mc_2.dat};
}
\node[v3panel, anchor=north west] at (rel axis cs:0.01,0.98) {(b) $S_1=+10$~dB};
\end{groupplot}
\end{tikzpicture}
\vspace{-4ex}
\caption{\gls{pfr}~\eqref{eq:pfr} of target~1 ($f_1=600$~Hz, $f_1T_f=0.192$) versus the number of frames $B$ in the scene of Table~\ref{tab:simul_params}. (a) Noise-limited, $S_1=-25$~dB. (b) Payload-limited, $S_1=+10$~dB. Energy-matched pilots: $K=\mathrm{round}(\mu_\gamma^2B)$ \gls{zc}-only frames. Lines: theory; markers: Monte Carlo.}
\label{fig:pfr}
\end{figure}
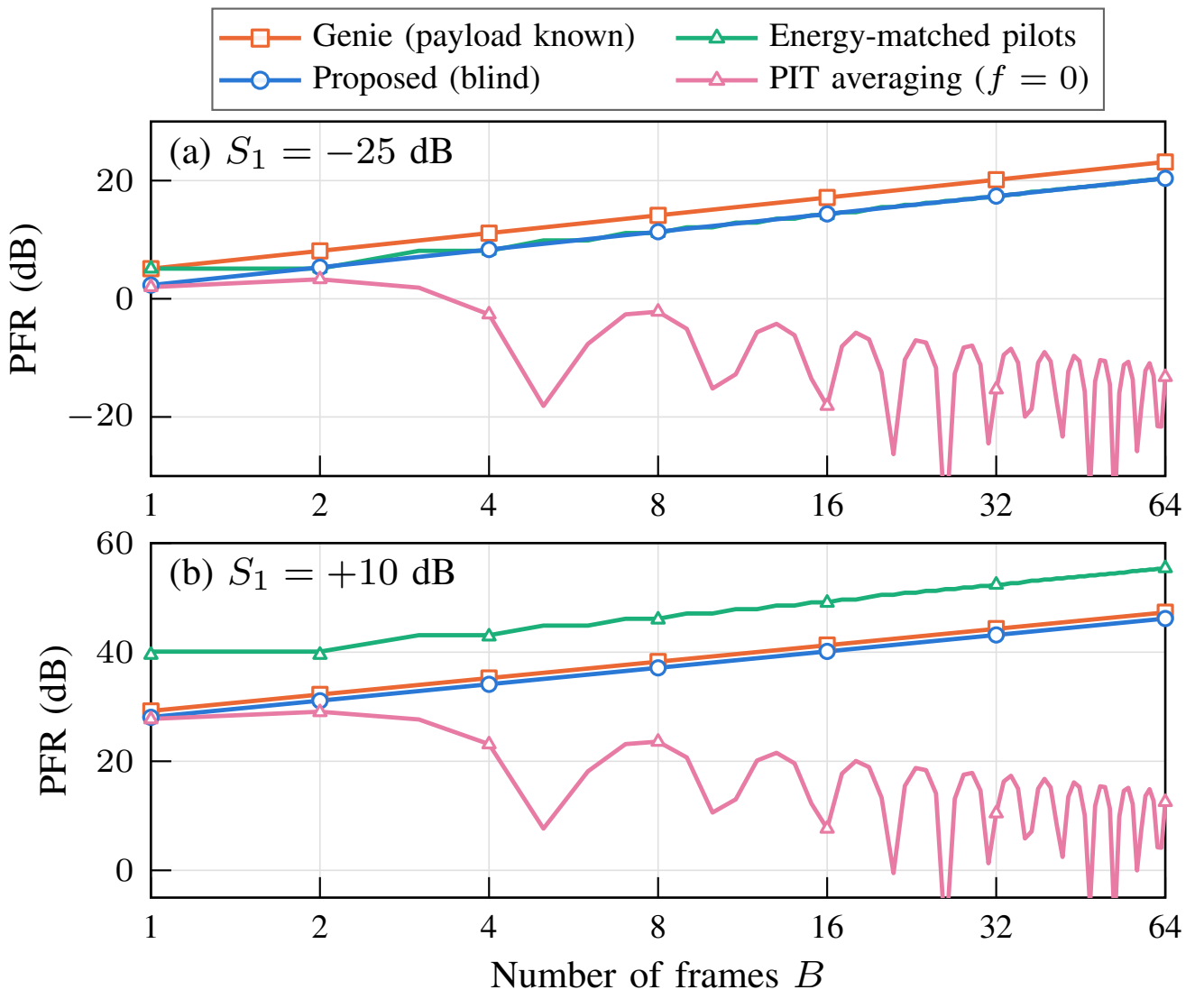

%Doppler alignment keeps the full $10\log_{10}B$ integration gain that \gls{pit} averaging forfeits (Fig.~\ref{fig:pfr}).
%At $S_1=-25$ and $+10$~dB, the noise lies $26$~dB above and $9$~dB below the payload term of~\eqref{eq:floor}, respectively.
%The \gls{pfr} of the proposed receiver grows from $2.3$ to $20.4$~dB and from $28.1$ to $46.2$~dB as $B$ increases from $1$ to $64$.
%The genie is $2.8$~dB better when noise-limited but only $1.1$~dB when payload-limited ($0.9$~dB asymptotically), where its own payload self-clutter dominates.
%Energy-matched pilots coincide with the proposed receiver when noise-limited, up to the rounding of $K$ at small $B$, and gain about $9$~dB at large $B$ when payload-limited.
%\gls{pit} averaging loses $20\log_{10}|D_B(f_1T_f)|^{-1}$ (plus $0.3$~dB of intra-frame loss), which is at least $20\log_{10}(B\sin\pi f_1T_f)$ ($19.2$~dB at $B=16$) and tends to complete cancellation as $f_1BT_f$ approaches a nonzero integer.
%At $B=16$, $f_1BT_f=3.07$ lies next to the third null, giving a $32.1$~dB loss, and the loss exceeds $50$~dB at $B=26$, $47$ and $52$.

Doppler alignment keeps the full $10\log_{10}B$ integration gain that \gls{pit} averaging forfeits (Fig.~\ref{fig:pfr}).
The genie is $2.8$~dB better when noise-limited but only $1.1$~dB when payload-limited ($0.9$~dB asymptotically), where its own payload self-clutter dominates.
Energy-matched pilots coincide with the proposed receiver when noise-limited, up to the rounding of $K$ at small $B$, and gain about $9$~dB at large $B$ when payload-limited.
For the simulated $600$~Hz target, \gls{pit} averaging loses about $32.5$~dB at $B=16$, including the intra-frame loss, because the uncompensated response lies near a slow-time null.

\subsection{Multi-Target Scene}

\begin{figure}[!t]
\centering
\includegraphics[width=\columnwidth]{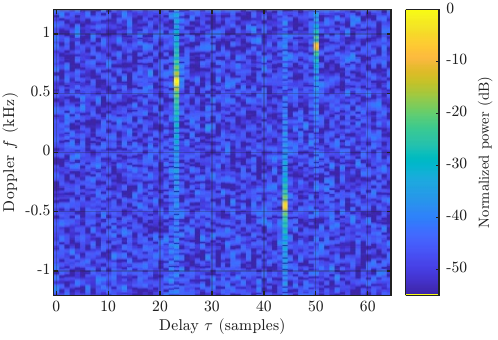}
\vspace{-4ex}
\caption{Payload-blind delay--Doppler map $|Z_B(\tau,f)|^2$ for $B=64$ frames in the scene of Table~\ref{tab:simul_params} at $S_1=20$~dB, normalized to the strongest peak.}
\label{fig:blind_map}
\end{figure}

A single payload-blind map resolves all three moving targets at their delay--Doppler coordinates (Fig.~\ref{fig:blind_map}).
The payload-induced floor averages about $-47$~dB relative to the strongest peak, as predicted by~\eqref{eq:floor}, with individual cells reaching about $-36$~dB.
In the same scene, averaging at $f=0$ as in \gls{pit} loses $33$--$35$~dB on all three targets, as given by~\eqref{eq:pfr}.
This high \gls{snr} exposes the payload floor; detection is treated next.

\subsection{Detection Performance}

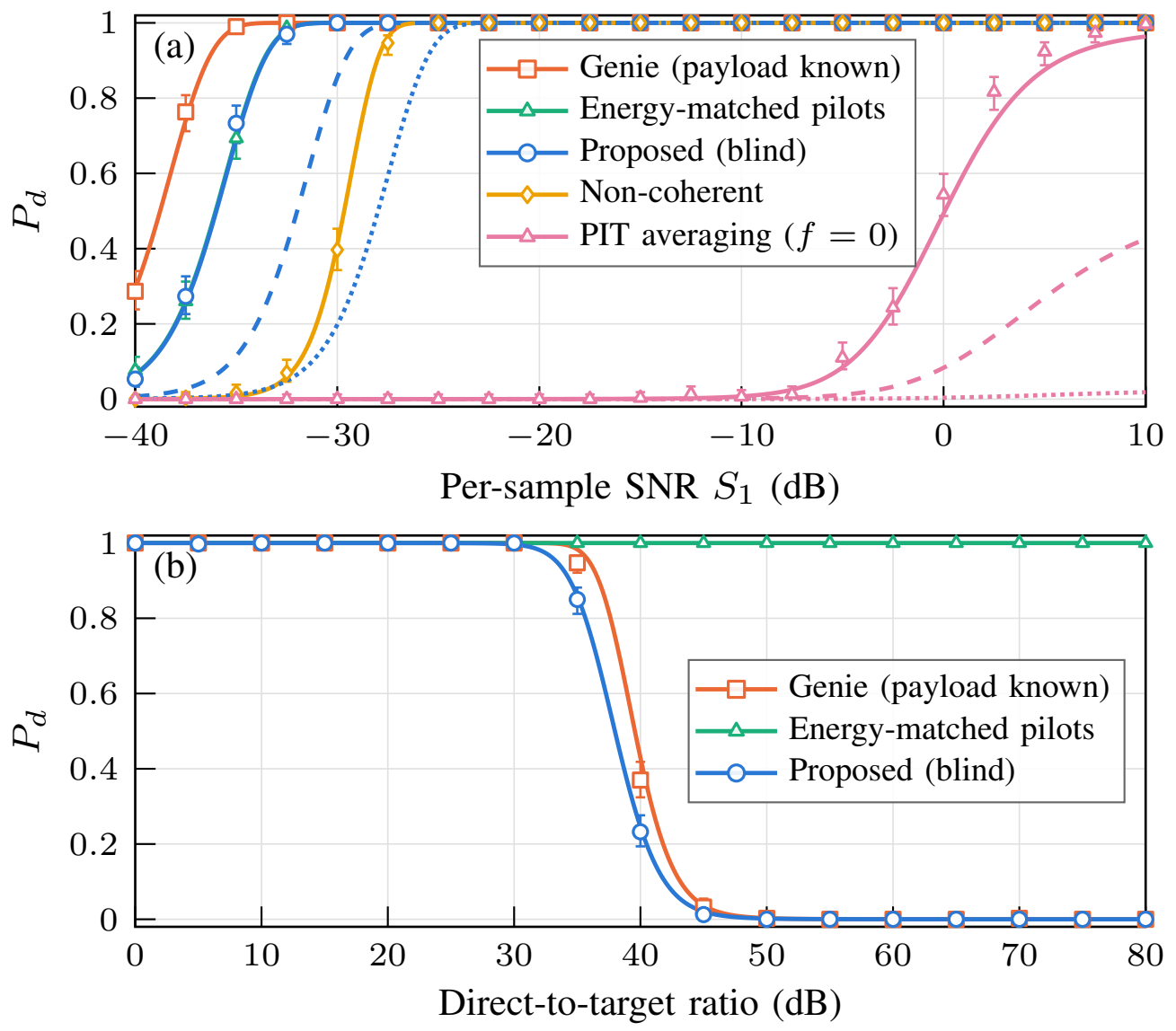
\begin{figure}[!t]
\centering
% Fig. 5: detection probability. Lines: theory (data/pd_th.dat, data/dp_th.dat); markers: Monte Carlo with
% Wilson 95% intervals (data/pd_mc.dat, data/dp_mc.dat). Data written by matlab_v3/export_tikz_data_v3.m.
\begin{tikzpicture}
\begin{groupplot}[group style={group size=1 by 2, vertical sep=27pt}, v3axis, height=82pt,
  ymin=-0.02, ymax=1.02, ytick={0,0.2,...,1}, ylabel={$P_d$},
  yticklabel style={/pgf/number format/.cd, fixed, precision=1}]
% (a) Pd versus S1; solid/dashed/dotted: targets 1/2/3
\nextgroupplot[xmin=-40, xmax=10, xtick={-40,-30,...,10}, xlabel={Per-sample SNR $S_1$ (dB)},
  legend style={at={(0.34,0.94)}, anchor=north west}]
\addlegendimage{v3leg=mgenie}\addlegendentry{Genie (payload known)}
\addlegendimage{v3leg=mpilot}\addlegendentry{Energy-matched pilots}
\addlegendimage{v3leg=mblind}\addlegendentry{Proposed (blind)}
\addlegendimage{v3leg=mnoncoh}\addlegendentry{Non-coherent}
\addlegendimage{v3leg=mpit}\addlegendentry{PIT averaging ($f=0$)}
\pgfplotsinvokeforeach{genie,pilot,blind,noncoh,pit}{
  \addplot[v3th=m#1] table[x=snr, y=#11] {figures_v3/data/pd_th.dat};
  \addplot[v3mc=m#1] table[x=snr, y=#1, y error minus=#1m, y error plus=#1p] {figures_v3/data/pd_mc.dat};
}
\pgfplotsinvokeforeach{blind,pit}{
  \addplot[v3th=m#1, dashed] table[x=snr, y=#12] {figures_v3/data/pd_th.dat};
  \addplot[v3th=m#1, densely dotted] table[x=snr, y=#13] {figures_v3/data/pd_th.dat};
}
\node[v3panel, anchor=north west] at (rel axis cs:0.01,0.98) {(a)};
% (b) Pd of target 1 versus the direct-to-target ratio, S1 = -30 dB
\nextgroupplot[xmin=0, xmax=80, xtick={0,10,...,80}, xlabel={Direct-to-target ratio (dB)},
  legend style={at={(0.98,0.5)}, anchor=east}]
\addlegendimage{v3leg=mgenie}\addlegendentry{Genie (payload known)}
\addlegendimage{v3leg=mpilot}\addlegendentry{Energy-matched pilots}
\addlegendimage{v3leg=mblind}\addlegendentry{Proposed (blind)}
\pgfplotsinvokeforeach{genie,pilot,blind}{
  \addplot[v3th=m#1] table[x=dtr, y=#1] {figures_v3/data/dp_th.dat};
  \addplot[v3mc=m#1] table[x=dtr, y=#1, y error minus=#1m, y error plus=#1p] {figures_v3/data/dp_mc.dat};
}
\node[v3panel, anchor=north west] at (rel axis cs:0.01,0.98) {(b)};
\end{groupplot}
\end{tikzpicture}
\vspace{-4ex}
\caption{Detection probability at the known target cell for a per-cell $P_{fa}=10^{-4}$ and $B=64$ in the scene of Table~\ref{tab:simul_params}. (a) $P_d$ versus $S_1$ (solid, dashed, dotted: targets~1, 2, 3; targets~2 and~3 for the proposed receiver and \gls{pit} only). (b) $P_d$ of target~1 versus the \gls{dtr} at $S_1=-30$~dB; the pilots ideally subtract the known direct-path \gls{zc} sequence. Lines: theory~\eqref{eq:pd} (chi-square for non-coherent integration); markers: Monte Carlo.}
\label{fig:pd}
\end{figure}

%For square-law detection, the threshold is the empirical $(1-P_{fa})$-quantile of the off-target cells pooled over all trials, i.e., the clairvoyant threshold that cell-averaging constant false-alarm rate detection approaches for a large window~\cite{richards2022}, and $P_d$ is evaluated at each target's known delay--Doppler cell.
%We also include non-coherent integration $\sum_b|c_b(\tau)|^2$ of the per-frame correlations $c_b(\tau)${\color{red}where $c_b(\tau)$ is defined? Isn't the \gls{pit} case the non-coherent one?}:, i.e.,~\eqref{eq:blind_combined} for frame $b$ alone with $f=0$, which needs no Doppler search and follows the chi-square counterpart of~\eqref{eq:pd}~\cite{richards2022}.
For square-law detection, the threshold is the empirical $(1-P_{fa})$-quantile of off-target cells pooled over all trials, and $P_d$ is evaluated at each target's known delay--Doppler cell. Equivalence to a large-window cell-averaging constant false-alarm rate threshold requires a homogeneous exponential null model~\cite{richards2022}.
The non-coherent and \gls{pit} baselines use the per-frame correlations in~\eqref{eq:frame_corr}.
The $K$ pilot frames lead the interval and have the coarser Doppler resolution $1/(KT_f)$.

Fig.~\ref{fig:pd}(a) shows that the proposed receiver reaches $P_d=0.9$ at $S_1=-33.6$, $-29.6$ and $-25.6$~dB for the three targets, so targets~2 and~3, which are $4$ and $8$~dB weaker, need exactly $4$ and $8$~dB more \gls{snr}, as predicted by~\eqref{eq:pfr}.
The genie is $2.8$~dB better, and energy-matched pilots perform within $0.2$~dB of the proposed receiver but spend $53\%$ of the frames on training.
Non-coherent integration, the payload-blind alternative without a Doppler search, needs $5.6$~dB more \gls{snr} for target~1 ($5.5$--$6$~dB over the three targets), which is the gain of Doppler alignment.
\gls{pit} averaging reaches $P_d=0.9$ for target~1 only at $S_1\approx4.5$~dB in simulation ($5.8$~dB by~\eqref{eq:pd}, conservatively) and attains only $P_d\lesssim0.45$ and $0.02$ for targets~2 and~3 at $S_1=10$~dB.

Fig.~\ref{fig:pd}(b) adds a direct path at $S_1=-30$~dB.
With $\Gamma_{\mathrm{req}}=11.75$~dB for $P_d=0.9$,~\eqref{eq:dtr} gives $\mathrm{DTR}_{\max}=37$~dB for $B=64$, which the noise reduces to $34$~dB at this \gls{snr}.
A conservative map-wide false-alarm probability of $10^{-2}$ over all $N\times2B$ cells raises $\Gamma_{\mathrm{req}}$ to $13.8$~dB, shifting the noise-limited curves of Fig.~\ref{fig:pd}(a) by about $2$~dB and lowering $\mathrm{DTR}_{\max}$ to $35$~dB, or about $30$~dB at $S_1=-30$~dB.
Energy-matched pilots are unaffected: in their training frames, the direct path is a known \gls{zc} sequence, removed here by ideal per-frame least-squares subtraction; in practice, the cancellation depth bounds their tolerance.
For a baseline $L=1$~km and a target with radar cross section $\sigma_{\mathrm{t}}=10$~m$^2$ at $R_T=R_R=1.4$~km from both nodes, $\mathrm{DTR}=4\pi R_T^2R_R^2/(\sigma_{\mathrm{t}}L^2)\approx67$~dB with isotropic antennas, or about $47$~dB with $20$~dB of receive-beam isolation, i.e., $13$--$33$~dB above the tolerance at $B=64$, which grows by $3$~dB per doubling of $NB$; longer integration, spatial isolation or direct-path suppression must close this gap.

These results assume on-grid Doppler shifts and constant target parameters.
The slow-time search can be implemented by a $2B$-point \gls{fft} across frames, whose worst-case straddle loss is $0.9$~dB ($3.9$~dB without zero-padding)~\cite{richards2022}, as confirmed by simulation; omitting the intra-frame compensation of~\eqref{eq:blind} would add up to $2.4$~dB at the edges of the unambiguous span $\pm1/(2T_f)=\pm1.56$~kHz, and the parameters must remain constant over $BT_f\approx20$~ms.

\section{Conclusion}
\label{sec:conclusion}

We showed that CE-OFDM with a phase-injected chirp, here a Zadoff--Chu sequence, supports payload-blind bistatic delay--Doppler sensing without dedicated sensing frames or an explicit pilot power split: the residual carrier becomes an implicit superimposed pilot, which a receive-only node integrates coherently across Doppler-aligned frames. 
Closed-form approximations characterize the payload-induced floor, PFR and known-cell detection probability.
For $\gamma=0.8$, the PFR gap relative to a payload-aware matched filter without interference cancellation is $0.9$~dB when payload-limited and $2.8$~dB when noise-limited.
At low SNR, the scheme matches a dedicated-pilot benchmark of equal reference energy without its $53\%$ frame overhead.
Doppler alignment saves $5.5$--$6$~dB over non-coherent integration, whereas PIT averaging loses about $32$~dB at $B=16$ for the simulated target.
Under the adopted interference model, the direct path limits payload-blind sensing to a predictable direct-to-target ratio that grows with the number of coherently integrated frames.
Given frame timing and the known chirp, the sensing receiver can use a $2B$-point FFT across frames, followed by Doppler-dependent fast-time compensation and chirp correlation, without payload demodulation.
The modulation index controls the reference fraction $e^{-\gamma^2}$ and the payload phase excursion.
Direct-path suppression, band-matched LFM and NLFM chirps for oversampled CE-OFDM, channel estimation and multistatic sensing are open themes for future work.

\end{document}